\documentclass[12pt]{iopart}
\usepackage{iopams}  

\usepackage{graphicx}

\def\ket#1{\vert#1\rangle}
\def\bra#1{\langle#1\vert}
\def\ip#1#2{\langle#1\vert#2\rangle}
\def\me#1#2#3{\langle#1\vert#2\vert#3\rangle}

\def\ph#1{\phantom{#1}}

\def\nn{\nonumber\\}
\def\beq{\begin{equation}}
\def\eeq{\end{equation}}

\def\i{{\bi i}}
\def\j{{\bi j}}
\def\r{{\bi r}}
\def\v{{\bi v}}
\def\k{{\bi k}}
\def\R{{\bi R}}

\def\O{{\boldsymbol{\it 0}}}
\def\eb{{\boldsymbol{\mathcal E}}}
\def\e{\mathcal E}
\def\mb{{\bi M}}
\def\ab{{\bi a}}
\def\bb{{\bi b}}
\def\cb{{\bi c}}
\def\lc{M^{\rm LC}}
\def\lcb{{\bi M}^{\rm LC}}
\def\ic{M^{\rm IC}}
\def\icb{{\bi M}^{\rm IC}}
\def\icO{M^{{\rm IC},0}}
\def\icOb{{\bi M}^{{\rm IC},0}}
\def\ice{M^{{\rm IC},\eb}}
\def\iceb{{\bi M}^{{\rm IC},\eb}}
\def\lct{{\widetilde M}^{\rm LC}}
\def\ict{\widetilde{M}^{\rm IC}}
\def\lctb{{\widetilde \mb}^{\rm LC}}
\def\ictb{\widetilde{\mb}^{\rm IC}}
\def\tri{\mathrm{Im\,}\tr}
\def\Tri{\mathrm{Im\,}\Tr}
\def\re{\mathrm{Re}}
\def\im{\mathrm{Im}}
\def\hk{H^0_\k}
\def\h0{H^0}
\def\ch{\mathcal H}

\def\mcs{M^{\rm CS}}
\def\mcsb{{\bi M}^{\rm CS}}
\def\alc{\widetilde{\alpha}^{\rm LC}}
\def\aic{\widetilde{\alpha}^{\rm IC}}
\def\acs{\alpha^{\rm CS}}
\def\tcs{\theta^{\rm CS}}
\def\tkubo{\theta^{\rm Kubo}}

\def\wt#1{\widetilde{#1}}

\def\part{\wt{\partial}}

\def\partb{\part_b}
\def\partc{\part_c}

\def\para{\partial_a}
\def\parb{\partial_b}
\def\parc{\partial_c}

\begin{document}

\title{Theory of orbital magnetoelectric response}

\author{Andrei Malashevich$^1$, Ivo Souza$^1$, Sinisa Coh$^2$
and David Vanderbilt$^2$}

\address{$^1$ Department of Physics, 
University of California,
Berkeley, CA 94720-7300, USA}

\address{$^2$ Department of Physics \& Astronomy, 
Rutgers University,
Piscataway, NJ 08854-8019, USA}

\ead{andreim@berkeley.edu}

\begin{abstract}
  We extend the recently-developed theory of bulk orbital
  magnetization to finite electric fields, and use it to calculate the
  orbital magnetoelectric response of periodic insulators. Working in
  the independent-particle framework, we find that the finite-field
  orbital magnetization can be written as a sum of three
  gauge-invariant contributions, one of which has no counterpart at
  zero field. The extra contribution is collinear with and explicitly
  dependent on the electric field. The expression for the orbital
  magnetization is suitable for first-principles implementations,
  allowing to calculate the magnetoelectric response coefficients by
  numerical differentiation. Alternatively, perturbation-theory
  techniques may be used, and for that purpose we derive an expression
  directly for the linear magnetoelectric tensor by taking the first
  field-derivative analytically.  Two types of terms are obtained.
  One, the `Chern-Simons' term, depends only on the unperturbed
  occupied orbitals and is purely isotropic. The other, `Kubo' terms,
  involve the first-order change in the orbitals and give isotropic as
  well as anisotropic contributions to the response. In ordinary
  magnetoelectric insulators all terms are generally present,
  while in strong $Z_2$ topological insulators only the
  Chern-Simons term is allowed, and is quantized. In order to
  validate the theory we have calculated under periodic boundary
  conditions the linear magnetoelectric susceptibility for a 3-D
  tight-binding model of an ordinary magnetoelectric insulator, using
  both the finite-field and perturbation-theory expressions. The
  results are in excellent agreement with calculations on bounded
  samples.
\end{abstract}


\submitto{\NJP}
\maketitle

\section{Introduction}
\label{sec:intro}

%
In insulating materials in which both spatial inversion and
time-reversal symmetries are broken, a magnetic field ${\bi B}$ can
induce a first-order electric polarization ${\bi P}$, and conversely
an electric field $\eb$ can induce a first-order magnetization
$\mb$~\cite{odell70,fiebig05}.  This linear magnetoelectric (ME)
effect is described by the susceptibility tensor
\beq 
\label{eq:alpha} 
\alpha_{da}=\left.\frac{\partial P_d}{\partial
    B_a}\right|_{{\bi B}=0} =\left.\frac{\partial M_a}{\partial
    \e_d}\right|_{\eb=0}
\eeq
where indices label spatial directions.  This tensor can be divided
into a ``frozen-ion'' contribution that occurs even when the ionic
coordinates are fixed, and a ``lattice-mediated'' contribution
corresponding to the remainder.  Each of these two contributions can
be decomposed further according to whether the magnetic interaction is
associated with spins or orbital currents, giving four
contributions to $\alpha$ in total.

All of those contributions, except the frozen-ion orbital one, are
relatively straightforward to evaluate, at least in principle, and
{\it ab initio} calculations have started to appear. For example, the
lattice-mediated spin-magnetization response was calculated in
\cite{iniguez08} for Cr$_2$O$_3$ and in \cite{wojdel09} for BiFeO$_3$
(including the strain deformation effects that are present in the
latter), and calculations based on the converse approach (polarization
response to a Zeeman field) were recently
reported~\cite{delaney09}. One generally expects the lattice-mediated
couplings to be larger than the frozen-ion ones, and insofar as the
spin-orbit interaction can be treated perturbatively, interactions
involving spin magnetization are typically larger than the orbital
ones.  However, we shall see that there are situations in which the
spin-orbit interaction cannot be treated perturbatively, and in which
the frozen-ion orbital contribution is expected to be dominant.
Therefore, it is desirable to have a complete description which
accounts for all four contributions.

The frozen-ion orbital contribution is, in fact, the one part of the
ME susceptibility for which there is at present no satisfactory
theoretical or computational framework, although some progress towards
that goal was made in two recent works\cite{qi08,essin09}.  Following
Essin {\it et al.}~\cite{essin09} we refer to it as the ``orbital
magnetoelectric polarizability'' (OMP).  For the remainder of this
paper, we will focus exclusively on this contribution to
\eref{eq:alpha}, and shall denote it simply by $\alpha$. Accordingly,
the symbol $\mb$ will be used henceforth for the orbital component of
the magnetization.

The question we pose to ourselves is the following: what is the
quantum-mechanical expression for the tensor $\alpha$ of a generic
three-dimensional band insulator?  We note that the conventional
perturbation-theory expression for $\alpha$~\cite{raab2005,barron04}
does not apply to Bloch electrons,
as it involves matrix elements of unbounded operators. The proper
  expressions for ${\bi P}$~\cite{King-Smith} and ${\bi
    M}$~\cite{xiao05,timo05,ceresoli06,shi07} in periodic crystals
  have been derived, but so far only at ${\bi B}=0$ and $\eb=0$
  respectively.  The evaluation of equation \eref{eq:alpha} remains
  therefore an open problem.

Phenomenologically, the most general form of $\alpha$ is a $3\times 3$
matrix where all nine components are independent.  Dividing it into
traceless and isotropic parts, the latter is conveniently expressed
in terms of a single dimensionless parameter $\theta$ as
\beq
\label{eq:alpha-iso}
\alpha^\theta_{da}=\frac{\theta e^2}{2\pi hc}\delta_{da}. 
\eeq
The presence of an isotropic ME coupling is equivalent to the addition
of a term proportional to $\theta\eb\cdot{\bi B}$ to the
electromagnetic Lagrangian. Such a term describes ``axion
electrodynamics'' \cite{wilczek87} and \eref{eq:alpha-iso} may
therefore also be referred to as the ``axion OMP.''  The
electrodynamic effects of the axion field are elusive (in fact, the
very existence of $\alpha^\theta$ was debated until recently: see
\cite{raab97,hehl08} and references therein).  For example, in a
finite, static sample cut from a uniform ME medium those effects are
only felt at the surface\cite{wilczek87,obukhov05}.  In
  particular, $\alpha^\theta$ gives rise to a surface Hall
  effect~\cite{widom86}.

An essential feature of the axion theory is that a change of $\theta$
by $2\pi$ leaves the electrodynamics invariant~\cite{wilczek87}. The
profound implications for the ME response of materials were recognized
by Qi {\it et al.}~\cite{qi08}, and discussed further by Essin
  {\it et al.}~\cite{essin09}.  These authors showed that there is a
part of the isotropic OMP which remains ambiguous up to integer
multiples of $2\pi$ in the corresponding $\theta$ until the surface
termination of the sample is specified. For example, a change by
  $2\pi n$ occurs if the surface is modified by adsorbing a quantum
  anomalous Hall layer.  Hence this particular contribution to
$\theta$ can be formulated as a bulk quantity only modulo a quantum of
indeterminacy, in much the same way as the electric polarization ${\bi
  P}$~\cite{King-Smith,resta-review07}. A microscopic expression for
it was derived in the framework of single-particle band theory by
the above authors. It is given by the Brillouin-zone integral of
the Chern-Simons form~\cite{csforms} in $k$-space, which is a
multivalued global geometric invariant reminiscent of the Berry-phase
expression for ${\bi P}$~\cite{King-Smith}.  We denote henceforth this
``geometric'' contribution to the OMP as the Chern-Simons OMP (CSOMP).

A remarkable outcome of this analysis is the prediction~\cite{qi08} of
a purely isotropic ``topological ME effect,'' associated with the
CSOMP, in a newly-discovered class of time-reversal invariant
insulators known as $Z_2$ topological
insulators~\cite{physicstoday,hasan10,moore10}.  As a result of the
multivaluedness of $\theta$, the presence of time-reversal symmetry in
the bulk, which takes $\theta$ into $-\theta$, is consistent with two
solutions: $\theta=0$, corresponding to ordinary insulators, and
$\theta=\pi$, corresponding to strong $Z_2$ topological
insulators.\footnote{An analogous situation occurs in the theory of
  polarization: inversion symmetry, which takes ${\bf P}$ into $-{\bf
    P}$, allows for a nontrivial solution which does not include ${\bf
    P}=0$ in the ``lattice'' of values~\cite{resta-review07}.  An
  important difference is that while $\theta$ is a directly measurable
  response, only {\it changes} in ${\bf P}$ are detectable, so that
  the experimental implications of the nontrivial solution are less
  clear in this case.} The latter case is non-perturbative in the
spin-orbit interaction, and $\theta=\pi$ amounts to a rather large ME
susceptibility (in Gaussian units it is $1/4\pi$ times the fine
structure constant, or $\sim$6$\times$10$^{-4}$, to be compared with
$\sim$1$\times$10$^{-4}$ for the total ME response of Cr$_2$O$_3$ at
low temperature~\cite{wiegelmann94}).

It is not clear from these recent works, however, whether the
isotropic CSOMP constitutes the full OMP response of a generic
insulator.  It does appear to do so for the tight-binding model
studied in \cite{essin09}, whose ME response was correctly reproduced
by the Chern-Simons expression even when the parameters were tuned to
break time-reversal and inversion symmetries (i.e., for generic
$\theta$ not equal to 0 or $\pi$).  On the other hand, other
considerations seem to demand additional contributions.  For example,
it is not difficult to construct tight-binding models of molecular
crystals in which it is clear that the OMP cannot be purely isotropic.

In this work we derive, using rigorous quantum-mechanical arguments,
an expression for the OMP tensor $\alpha$ of band insulators, written
solely in terms of bulk quantities (the periodic Hamiltonian and
ground state Bloch wavefunctions, and their first-order change in an
electric field).  We restrict our derivation to non-interacting
Hamiltonians, as the essential physics we wish to describe occurs
already at the single-particle level.  We find that in crystals with
broken time-reversal and inversion symmetries there are, in addition
to the CSOMP term discussed in \cite{qi08,essin09}, extra terms which
generally contribute to both the trace and the traceless parts of
$\alpha$.

Our theoretical approach closely mimics one type of ME response
experiment: a finite electric field $\eb$ is applied to a bounded
sample, and the (orbital) magnetization is calculated in the presence
of the field.  Then the thermodynamic limit is taken at fixed field.
This key step in the derivation must be done carefully, so that
crucial surface contributions are not lost in the process, and
  here we follow the Wannier-based approach of
  references~\cite{timo05,ceresoli06}, adapted to $\eb\not=0$.  Finally the linear response
coefficient $\alpha_{da}=\partial M_a/\partial \e_d$ is extracted in
the limit that $\eb$ goes to zero.

In a concurrent work by Essin, Turner, Moore, and one of
  us~\cite{essin10} an alternative approach was taken, which is closer
  in spirit to the calculation in \cite{King-Smith} of the change in
  polarization as an integrated current: the adiabatic current induced
  in an infinite crystal by a change in its Hamiltonian in the
  presence of a magnetic field is computed, and then expressed as a
  total time derivative.  The two approaches are complementary and
  lead to the same expression for $\alpha$, illuminating it from
  different angles.

The paper is organized as follows. In
  section~\ref{sec:finite-field} we derive the bulk expression for
  ${\bf M}(\eb)$, and reorganize it into three gauge-invariant
  contributions, one of which yields directly the CSOMP response.
The gauge-invariant decomposition of $\mb(\eb)$ is done at first in
$k$-space for periodic crystals, and then also for bounded samples
working in real space.  In section~\ref{sec:linear-response} we derive
a $k$-space formula for the OMP tensor $\alpha$ by taking analytically
the field-derivative of ${\bi M}(\eb)$.  Numerical tests on a
tight-binding model of a ME insulator are presented at appropriate
places throughout the paper in order to validate the bulk expressions
for $\mb(\eb)$ and $\alpha$.  In \ref{app:model} we describe the
tight-binding model, as well as technical details on how the various
formulas are implemented on a $k$-point grid. \ref{app:derivation} and
\ref{app:band-sum-consistency} contain derivations of certain results
given in the main text.

\section{Orbital magnetization in finite electric field}
\label{sec:finite-field}

\subsection{Preliminaries}

The orbital magnetization $\mb$ is defined as the orbital moment per
unit volume,
\beq
\label{eq:M-finite}
\mb=-\frac{e}{2cV}\sum_i\,\bra{\psi_i}\r\times\v\ket{\psi_i}.
\eeq
Here $e>0$ is the magnitude of the electron charge, $V$ is the sample
volume, and $\ket{\psi_i}$ are the occupied eigenstates.  While this
expression can be directly implemented when using open boundary
conditions, the electronic structure of crystals is more conveniently
calculated and interpreted using periodic boundary conditions,
in order to take advantage of Bloch's theorem. This poses however
serious difficulties in dealing with the circulation operator
$\r\times\v$, because of the unbounded and nonperiodic nature of the
position operator $\r$.  These subtle issues were fully resolved only
recently, with the derivation of a bulk expression for $\mb$ directly
in terms of the extended Bloch
states~\cite{xiao05,timo05,ceresoli06,shi07}.

In previous derivations the crystal was taken to be under shorted
electrical boundary conditions.  We shall extend the derivation given
in \cite{timo05,ceresoli06} to the case where a static homogeneous
electric field $\eb$ is present, so that the full Hamiltonian reads
\beq
\label{eq:ham}
\ch=\ch^0+e\eb\cdot\r.
\eeq
The derivation, carried out for an insulator with $N$ valence bands
within the independent-particle approximation, involves transforming
the set of occupied eigenstates $\ket{\psi_i}$ of $\ch$ into a set of
Wannier-type (i.e., localized and orthonormal) orbitals $\ket{w_i}$
and expressing $\mb(\eb)$ in the Wannier representation. This is done
at first for a finite sample cut from a periodic crystal, and
eventually the thermodynamic limit is taken at fixed field.

Before continuing, two remarks are in order. First, the assumption
that it is possible to construct well-localized Wannier functions
(WFs) spanning the valence bands is only valid if the Chern invariants
of the valence bandstructure vanish identically~\cite{timo06}. This
requirement is satisfied by normal band insulators as well as by $Z_2$
topological insulators, but not by quantum anomalous Hall
insulators~\cite{haldane88}, which thus far remain
hypothetical. Second, because of Zener tunnelling, an insulating
crystal does not have a well-defined ground state in a finite electric
field.  Nonetheless, upon slowly ramping up the field to the desired
value, the electron system remains in a quasistationary state which
is, for all practical purposes, indistinguishable from a truly
stationary state. This is the state we shall consider in the ensuing
derivation.  As discussed in \cite{souza02,souza04}, it is Wannier-
and Bloch-representable, even though the Hamiltonian \eref{eq:ham} is
not lattice-periodic.

\subsection{$k$-space expression}
\label{sec:finite-field-k}

Our derivation of a $k$-space (bulk) expression for $\mb(\eb)$ is carried out
mostly in real space, using a Wannier representation. It is only in
the last step that we switch to reciprocal space, by expressing the
crystalline WFs $\ket{\R n}$ in terms of the cell-periodic Bloch
functions $\ket{u_{n\k}}$ via \cite{mv97}
\beq 
\ket{\R n}=V_{\rm c}\int [\rmd k]\rme^{\rmi\k\cdot(\r-\R)}\ket{u_{n\k}},
\eeq 
where $\R$ is a lattice vector, $V_{\rm c}$ is the unit-cell volume,
$[\rmd k]\equiv \rmd^3k/(2\pi)^3$, and the integral is over the first
Brillouin zone.

We begin with a finite sample immersed in a field $\eb$, divide it up
into an interior region and a surface region, and assign each WF to
either one. The boundary between the two regions is chosen in such a
way that the fractional volume of the surface region goes to zero as
$V\rightarrow\infty$, but deep enough that WFs near the boundary are
bulk-like. Following \cite{timo05,ceresoli06}, equation
\eref{eq:M-finite} for the orbital magnetization can then be rewritten
as an interior contribution plus a surface contribution, denoted
respectively as the ``local circulation'' (LC) and the ``itinerant
circulation'' (IC).  Remarkably, in the thermodynamic limit {\it both}
can be expressed solely in terms of the interior-region crystalline
WFs, or equivalently, in terms of the bulk Bloch functions, as shown
in the above references at $\eb=0$ and below for $\eb\not=
0$. Specifically, we shall show that \numparts
\beq
\label{eq:M-tot-a}
\mb=\lcb+\icOb+\iceb,
\eeq
where
\beq
\label{eq:M-lc-bulk}
\lc_a=-\gamma\epsilon_{abc}\im
\sum_{n}^N\,\int [\rmd k]\bra{\partial_b u_{n\k}}
\hk\ket{\partial_c u_{n\k}}
\eeq
is the contribution from the interior WFs,
\beq
\label{eq:M-ic0-bulk}
M^{{\rm IC},0}_a=-\gamma\epsilon_{abc}\im
\sum_{nm}^N\,\int [\rmd k]\ip{\partial_b u_{n\k}}
{\partial_c u_{m\k}}H^0_{mn\k}
\eeq
is the part of the surface contribution 
coming from the zero-field Hamiltonian, and
\beq
\label{eq:M-ice-bulk}
M^{{\rm IC},\eb}_a=-\gamma\epsilon_{abc}\im
\sum_{nm}^N\,\int [\rmd k]\ip{\partial_b u_{n\k}}
{\partial_c u_{m\k}}e\eb\cdot{\bi A}_{mn\k}
\eeq
\endnumparts
is the part of the surface contribution coming from the electric field
term in the Hamiltonian \eref{eq:ham}. In the above
expressions 
$\gamma=-e/(2\hbar c)$,
\beq
\label{eq:hk}
\hk=\rme^{-\rmi\k\cdot\r}\ch^0 \rme^{\rmi\k\cdot\r},
\eeq
\beq
\label{eq:h0}
H^0_{mn\k}=\bra{u_{m\k}}\hk\ket{u_{n\k}},
\eeq 
and ${\bf A}_{mn\k}$ is the Berry connection matrix defined in
equation \eref{eq:Amnb} below.

Having stated the result we now present the derivation, starting with
the interior contribution $\lcb$.  Using $[r_i,r_j]=0$, the velocity
operator $\v=(\rmi/\hbar)[\ch,\r]$ becomes $(\rmi/\hbar)[\ch^0,\r]$, so that the
circulation operator $\r\times\v$ is unaffected by the electric
field. It immediately follows that the local circulation part $\lcb$
is given in terms of the field-polarized states $\ket{u_{n\k}}$ by
the same expression, equation~\eref{eq:M-lc-bulk}, as was derived in
\cite{ceresoli06} for the zero-field case.

Consider now the contribution 
$\icb=\icOb+\iceb$ from the surface WFs $\ket{w_s}$.
For large samples it takes the form~\cite{ceresoli06}
\beq
\icb=-\frac{e}{2cN_{\rm c}V_{\rm c}}\sum_s^{\rm surf}\,\r_s\times\v_s,
\eeq
where $N_{\rm c}$ is the number of crystal cells of volume $V_{\rm c}$,
$\r_s=\bra{w_s}\r\ket{w_s}$, and
\beq
\label{eq:v_s}
\v_s=\bra{w_s}\v\ket{w_s}=\frac{2}{\hbar}\im\bra{w_s}\r \ch\ket{w_s}.
\eeq
Note that $\ch\ket{w_s}$ already belongs to the occupied manifold
spanned by $P=\sum_j^{\rm occ}\,\ket{w_j}\bra{w_j}$, since we
assume a (quasi)stationary state.  Thus we can insert a $P$
between $\r$ and $\ch$ above, and using \eref{eq:ham}
we obtain
\beq
\v_s=
\sum_j^N\,\left(
  \v_{\langle js\rangle}^0+\v_{\langle js\rangle}^\eb
  \right),
\eeq
where $\v_{\langle js\rangle}^0=(2/\hbar)\im[\r_{sj}\ch^0_{js}]$ is the
same as in \cite{timo05,ceresoli06} 
and $\v_{\langle js\rangle}^\eb=(2e/\hbar)\im[\r_{sj}(\r_{js}\cdot\eb)]$
is a new term.  

The reasoning~\cite{timo05,ceresoli06} by which $\icb$ can be recast in
terms of the bulk WFs $\ket{\R n}$ relies on the exponential
localization of the WFs and on certain properties of $\v_{\langle js\rangle}^0$
(antisymmetry under $j\leftrightarrow s$ and invariance under lattice
translations deep inside the crystallite) which are shared by
$\v_{\langle js\rangle}^\eb$. Hence we can follow similar steps
as in those works, arriving at
\beq
\label{eq:ice_a}
\ice_a=\frac{e}{4cV_{\rm c}}\epsilon_{abc}
\sum_\R\sum_{mn}^N\, v^{\eb}_{\langle
\O m,\R n\rangle,b}R_c,
\eeq
and similarly for $\icO_a$ with $v^0$ substituting for $v^{\eb}$.  The
latter is identical to the expression for $\ic_a$ valid at
$\eb=0$~\cite{timo05,ceresoli06}, and upon converting to $k$-space
becomes \eref{eq:M-ic0-bulk}.

Let us now turn to $\ice_a$ and write \eref{eq:ice_a} as
$(e^2/2c\hbar V_{\rm c})\epsilon_{abc}\e_d\im W_{bd,c}$ where
\beq 
\label{eq:W_bdc}
W_{bd,c}=\sum_\R\sum_{mn}^N\,\bra{\R n}r_b\ket{\O m}
\bra{\O m}r_d\ket{\R n}R_c.
\eeq
In order to recast this expression as a $k$-space integral it is
useful to introduce the $N\times N$ Berry connection matrix
\beq
\label{eq:Amnb}
A_{mn\k,b}=\rmi\bra{u_{m\k}}\partial_b u_{n\k}\rangle=A_{nm\k,b}^*,
\eeq
where $\partial_b\equiv\partial/\partial k_b$.  
It satisfies the
relation~\cite{mv97,blount1962}
\beq
\label{eq:A-w1}
\bra{\R n}r_b\ket{\O m}=V_c\int[\rmd k]A_{nm\k,b}\rme^{\rmi\k\cdot\R}.
\eeq
We also need
\beq
\label{eq:A-w2}
R_c\bra{\R n}r_d\ket{\O m}=\rmi V_c\int [\rmd k](\partial_c A_{nm\k,d})\rme^{\rmi\k\cdot\R},
\eeq
which follows from \eref{eq:A-w1}.
Using these two relations, \eref{eq:W_bdc} becomes
\beq
W_{bd,c}=\rmi V_c\sum_{mn}^N\int[\rmd k]A_{mn\k,d}\partial_c A_{nm\k,b},
\eeq
and we arrive at \eref{eq:M-ice-bulk}.

The sum of equations~\eref{eq:M-lc-bulk}--\eref{eq:M-ice-bulk} gives the
desired $k$-space expression for $\mb(\eb)$.
In the limit $\eb\rightarrow 0$ the term $\iceb$ vanishes, and
equation (31) of \cite{ceresoli06} is recovered.

\begin{figure}
\centering\includegraphics{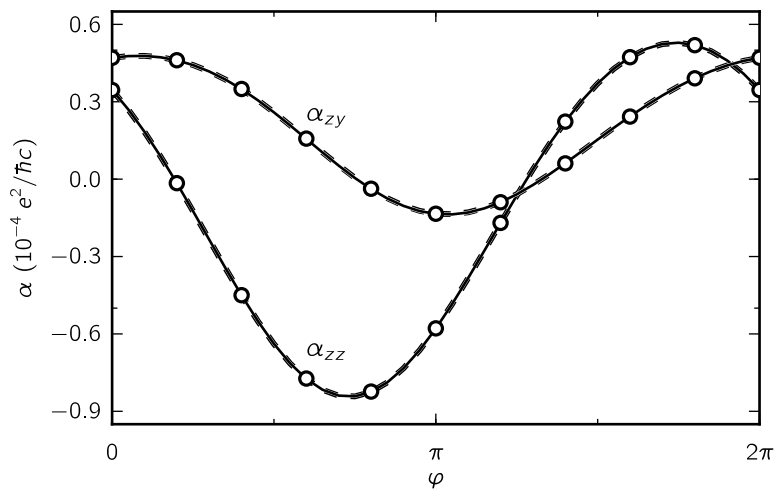}
\caption{The $zz$ and $zy$ components of the OMP tensor
  $\alpha$ of the tight-binding model described in
  \ref{app:model}, as a function of the parameter $\varphi$.
  The two lower bands are treated as occupied. Solid line:
  extrapolation from finite-size samples using numerical
  differentiation of the finite-field magnetization calculated from
  \eref{eq:M-finite}.  Dashed line: numerical differentiation of
  the finite-field magnetization calculated using
  \eref{eq:M-lc-bulk}--\eref{eq:M-ice-bulk} discretized on a
  $k$-space grid.  Open circles: linear-response calculation in
  $k$-space using discretized versions of
  \eref{eq:theta-cs}--\eref{eq:alpha-ic}.  }
\label{fig:al_zz}
\end{figure}

We have implemented \eref{eq:M-lc-bulk}--\eref{eq:M-ice-bulk} for the
tight-binding model of \ref{app:model}. Since for small electric
fields $\mb(\eb)$ differs only slightly from $\mb(0)$, in order to
observe the effect of the electric field we consider differences in
magnetization rather than the absolute magnetization.  Therefore, in
all our numerical tests we evaluated the OMP tensor
$\alpha_{da}$. With the help of
\eref{eq:M-lc-bulk}--\eref{eq:M-ice-bulk} we calculated it as $\Delta
M_a/\Delta \e_d$, using small fields $\e_d=\pm 0.01$.  We then
repeated the calculation on finite samples cut from the bulk crystal,
using \eref{eq:M-finite} in place of
\eref{eq:M-lc-bulk}--\eref{eq:M-ice-bulk}. Figure~\ref{fig:al_zz}
shows the value of the $zz$ and $zy$ components of $\alpha$ plotted as
a function of the parameter $\varphi$, the phase of one of the complex
hopping amplitudes (see \ref{app:model} for details). The very precise
agreement between the solid and dashed lines confirms the correctness
of the $k$-space formula.  The same level of agreement was found for
the other components of $\alpha$.

\subsection{Gauge-invariant decomposition}

\subsubsection{Periodic crystals}
\label{sec:finite-field-k-inv}

\Eref{eq:M-tot-a} for $\mb(\eb)$ is valid in an arbitrary
gauge, that is, the sum of its three terms given by
\eref{eq:M-lc-bulk}--\eref{eq:M-ice-bulk} --~but not each term
individually~-- remains invariant under a unitary transformation
\beq
\label{eq:gauge}
\ket{u_{n\k}}\rightarrow \sum_m^N\,\ket{u_{m\k}}U_{mn\k}
\eeq
among the valence-band states at each $\k$.  In order to make the
gauge invariance of \eref{eq:M-tot-a} manifest, it is convenient to
first manipulate it into a different form, given in terms of certain
canonical objects which we now define.  We begin by introducing the
covariant $k$-derivative of a valence state~\cite{souza04},
\beq
\label{eq:cov-der}
 \ket{\partb u_{n\k}} =Q_\k\,\ket{\parb u_{n\k}},
\eeq
where $Q_\k=1-P_\k$ and 
\beq
\label{eq:proj}
P_\k=\sum_{j=1}^N\ket{u_{j\k}}\bra{u_{j\k}}.
\eeq
The covariant and ordinary derivatives are related by
\beq
\label{eq:cov_deriv}
\ket{\parb u_{n\k}} = \ket{\partb u_{n\k}} 
-\rmi \sum_m^N A_{mn\k,b}\ket{u_{m\k}}.
\eeq
The generalized metric-curvature tensor is~\cite{mv97}
\beq
\label{eq:metric-curv}
F_{nm\k,bc}=\ip{\partb u_{n\k}}{\partc u_{m\k}} = 
F_{mn\k,cb}^*.
\eeq
Viewed as an $N\times N$ matrix over the band indices, $F$ is
gauge-covariant, changing as
\beq
\label{eq:gauge-covariant}
F_{nm\k,bc}\rightarrow \left(U_\k^\dagger F_{\k,bc} U_\k\right)_{nm}
\eeq
under the transformation \eref{eq:gauge}.  We also note the relation
\beq
\label{eq:F-ident}
\ip{\parb u_{n\k}}{\parc u_{m\k}} = F_{nm\k,bc}+(A_{\k,b}A_{\k,c})_{nm}.
\eeq
We shall make use of two more gauge-covariant objects,
\beq
H^0_{nm\k,b}=\rmi\me{u_{n\k}}{\hk}{\partb u_{m\k}}
\label{eq:Xa}
\eeq
and
\beq
\label{eq:Xab}
H^0_{nm\k,bc}=\me{\partb u_{n\k}}{\hk}{\partc u_{m\k}},
\eeq
which enter the relation
\beq 
\label{eq:Xnmab-ident}
\fl\me{\parb u_{n\k}}\hk{\parc u_{m\k}}= 
H^0_{nm\k,bc}
+\left[A_{\k,b} H^0_{\k,c} + 
\left( H^0_{\k,b}\right)^\dagger A_{\k,c}+
A_{\k,b}\hk A_{\k,c}\right]_{nm}.
\eeq

Coming back to equations \eref{eq:M-tot-a}--\eref{eq:M-ice-bulk}, for
$\lc_a$ we use \eref{eq:Xnmab-ident} and for $\ic_a$ we use
\eref{eq:F-ident}, leading to
\beq
\fl M_a=-\gamma\epsilon_{abc}\int [\rmd k]\,
  \tri\Big[
  H^0_{bc} + 2A_bH^0_c
  +H^0F_{bc}
 + e\e_dA_dF_{bc} +
  e\e_d A_d A_b A_c
  \Big],
\eeq
where ``tr'' denotes the electronic trace over the occupied valence
bands and we have dropped the subscript $\k$. The second term can be
rewritten using 
\beq
\label{eq:H0nmc}
H^0_{nm,c}=-e\e_dF_{nm,dc}.
\eeq
(To obtain this relation start from the generalized Schr\"odinger equation
satisfied by the valence states at $\eb\not=0$~\cite{nunes01},
\beq
H^0\ket{u_n}=\sum_m^N\,
(H^0_{mn}+e\eb\cdot{\bi A}_{mn})
\ket{u_m}-
\rmi e\e_d\ket{\partial_d u_n}, 
\eeq
and multiply through by $\bra{\partc u_m}$.)  

Let us define the quantities
\beq
\label{eq:M-lc-a}
\lct_a=-\gamma\epsilon_{abc}\int [\rmd k]\tri \left[ H^0_{bc}\right],
\eeq
\beq
\label{eq:M-ic-a}
\ict_a=-\gamma\epsilon_{abc}\int [\rmd k]\tri \left[ H^0F_{bc}\right],
\eeq
and
\beq
\label{eq:M-cs-a}
\mcs_a=-e\gamma\epsilon_{abc}\e_d\int [\rmd k]\tri
 \bigl[
2A_bF_{cd}+F_{bc}A_d
+A_bA_cA_d
\bigr].
\eeq
The total magnetization is given by their sum
\numparts
\beq
\label{eq:M-tot}
M_a=\lct_a+\ict_a+\mcs_a.
\eeq
Referring to \eref{eq:metric-curv} and \eref{eq:Xab} the first
two terms read, in a more conventional notation,
\beq
\label{eq:M-lc}
\lct_a=-\gamma\epsilon_{abc}\int [\rmd k]\sum_n^N\,\im
\bra{\partb u_{n\k}}\hk\ket{\partc u_{n\k}}
\eeq
and
\beq
\label{eq:M-ic}
\ict_a=-\gamma\epsilon_{abc}\int [\rmd k]\sum_{nm}^N\,\im
\left(
  H^0_{nm\k}\bra{\partb u_{m\k}}\partc u_{n\k}\rangle
\right).
\eeq
These are the only terms that remain in the limit $\eb\rightarrow 0$,
in agreement with equation~(43) of \cite{ceresoli06}.
At finite field they depend on $\eb$ implicitly via the wavefunctions.

We now show that the
term $\mcsb$, which gathers all the contributions with an
explicit dependence on $\eb$, can be recast as
\beq
\label{eq:M-cs}
\mcs_a=e\gamma\e_a\int [\rmd k]\epsilon_{ijk}\tr\left[A_i\partial_jA_k-
\frac{2\rmi}{3}A_iA_jA_k\right].
\eeq
\endnumparts
To do so it is convenient to introduce the Berry curvature tensor
\beq
\label{eq:omega}
\Omega_{nm,ab}=\rmi F_{nm,ab}-\rmi F_{nm,ba}=-\Omega_{nm,ba},
\eeq
where $F_{nm,ab}$ was defined in \eref{eq:metric-curv}.
A few lines of algebra show that
\beq
\label{eq:omega-b}
\Omega_{nm,ab} = \para A_{nm,b} - \parb A_{nm,a} -\rmi[A_a,A_b]_{nm}.
\eeq
In order to go from \eref{eq:M-cs-a} to \eref{eq:M-cs}, use
\eref{eq:omega} to write $\tri[F_{bc}A_d]$ as
$-\frac12\tr[A_d\Omega_{bc}]$ and $-2\tri[A_bF_{dc}]$ as
$\tr[A_b\Omega_{dc}]$, and then replace $\Omega_{nm,bc}$ in these
expressions with $\epsilon_{abc}\Omega_{nm,a}$, 
where $\Omega_{nm,a}=\frac12\epsilon_{abc}\Omega_{nm,bc}$ is the Berry
curvature tensor written in axial-vector form.  This leads to
\beq
\label{eq:M-cs-alt}
\mcs_a=
e\gamma\int [\rmd k]
\big(
\e_a\tr[\bOmega\cdot{\bi A}]-
\e_d\epsilon_{abc}\tri[A_bA_cA_d]
\big).
\eeq
The first term is parallel to the field, and can be rewritten with the
help of \eref{eq:omega-b}:
\beq
\label{eq:omega-A}
\tr[\bOmega\cdot{\bi A}]=\epsilon_{ijk}
\tr[A_i\partial_j A_k-\rmi A_iA_jA_k].
\eeq
While not immediately apparent, the second term in
\eref{eq:M-cs-alt} also points along the field. To see this,
write
\beq
\fl\sum_{bcd}\,
\e_d\epsilon_{abc}\tri[A_bA_cA_d]=
\e_a\sum_{bc}\,\epsilon_{abc}\tri[A_aA_bA_c]
+
\sum_{d\not= a}\sum_{bc}\,\epsilon_{abc}\tri[A_bA_cA_d],
\eeq
where we suspended momentarily the implied summation convention.  The
last term vanishes because the factor $\epsilon_{abc}$ forces $d\not=
a$ to equal either $b$ or $c$, producing terms such as
$\tri[A_bA_bA_c]$ which vanish identically as $A_b$ is
Hermitian. Rewriting $\e_a\sum_{bc}\,\epsilon_{abc}\tri[A_aA_bA_c]$ as
$(\e_a/3)\sum_{ijk}\epsilon_{ijk}\tri[A_iA_jA_k]$ and restoring the
summation convention, we arrive at \eref{eq:M-cs}.

Equations~\eref{eq:M-lc}--\eref{eq:M-cs}, which constitute the main
result of this section, are separately gauge-invariant.  For
$\wt{\mb}^{\rm LC}$ and $\wt{\mb}^{\rm IC}$ this is apparent already
from \eref{eq:M-lc-a} and \eref{eq:M-ic-a}, whose {\it integrands} are
gauge-invariant, being traces over gauge-covariant matrices.  In
contrast, equation~\eref{eq:M-cs} for $\mcsb$ only becomes invariant
after taking the integral on the right-hand-side over the entire
Brillouin zone (the integrand being familiar from differential
geometry as the Chern-Simons 3-form~\cite{nakahara,csforms}).

The Chern-Simons contribution \eref{eq:M-cs} has several remarkable
features: (i) as already noted, it is perfectly isotropic, remaining
parallel to $\eb$ for arbitrary orientations of $\eb$ relative to the
crystal axes; (ii) being isotropic, it vanishes in less than three
dimensions, which intuitively can be understood because already in two
dimensions polarization must be in the plane of the system and
magnetization must be out of the plane; (iii) for $N>1$ valence bands
it is a multivalued bulk quantity with a quantum of arbitrariness
$(e^2/hc)\e_a$, a fact that is connected with the possibility of a
cyclic adiabatic evolution that would change \eref{eq:theta-cs} below
for $\theta$ by $2\pi$~\cite{qi08}.

\begin{figure}
\centering\includegraphics{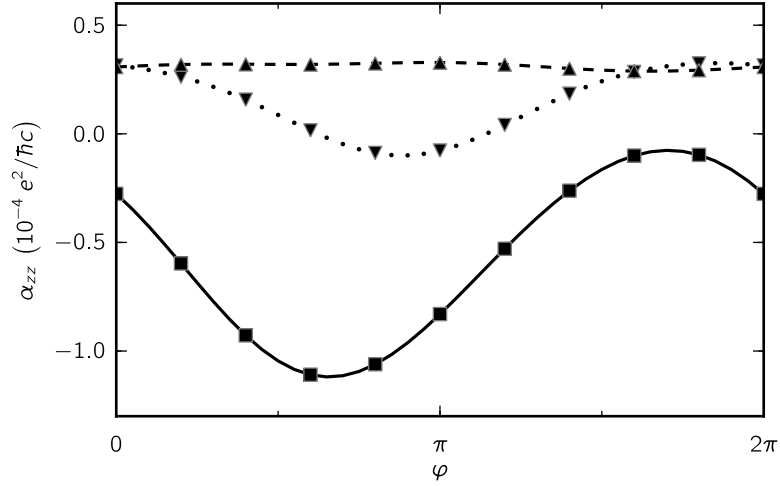}
\caption{Decomposition of the $\alpha_{zz}$ curve in
  figure~\ref{fig:al_zz} into the gauge-invariant contributions
  $\alc_{zz}$ (solid lines), $\aic_{zz}$ (dashed line), and
  $\acs_{zz}$ (dotted line), calculated in $k$-space using finite
  differences in $\eb$. Symbols denote the same contributions
  evaluated for bounded samples, also using finite differences.
}
\label{fig:al_zz_c}
\end{figure}

We have repeated the calculation of the OMP presented in
figure~\ref{fig:al_zz} using \eref{eq:M-lc}--\eref{eq:M-cs} instead of
\eref{eq:M-lc-bulk}--\eref{eq:M-ice-bulk}, finding excellent agreement
between them. The electric field derivative of the
decomposition~\eref{eq:M-tot} gives the corresponding decomposition of
the OMP tensor \eref{eq:alpha},
\beq
\label{eq:omp-tot}
\alpha=\alc+\aic+\acs,
\eeq
where each term is also gauge-invariant. The $zz$ components of these
terms are plotted separately in figure~\ref{fig:al_zz_c}.

\subsubsection{Finite samples}

It is natural to ask whether the gauge-invariant decomposition of the
orbital magnetization given in equation \eref{eq:M-tot} can be made
already for finite samples, before taking the thermodynamic limit and
switching to periodic boundary conditions. This has previously been
done in the case $\eb=0$, where $\mcsb=0$ and $\lctb$ and $\ictb$ take
the form~\cite{souza08}
\numparts
\beq
\label{eq:M-lcfin}
\lct_a=\frac{e}{2\hbar cV}\epsilon_{abc}\mathrm{Im\,}\Tr\,[Pr_bQ\ch^0Qr_c]
\eeq
and
\beq
\label{eq:M-icfin}
\ict_a=\frac{e}{2\hbar cV}\epsilon_{abc}\mathrm{Im\,Tr}\,[P\ch^0Pr_bQr_c].
\eeq
Here $P$ and $Q=1-P$ are the projection operators onto the occupied and 
empty subspaces, respectively, and
``Tr'' denotes the electronic trace over the entire Hilbert space.
These two expressions, which are manifestly gauge-invariant, remain
valid at finite field, reducing to \eref{eq:M-lc} and
\eref{eq:M-ic} in the thermodynamic limit.

We now complete this picture for $\eb\not= 0$ by showing that the
remaining contribution $\mcsb=\mb-\lctb-\ictb$ can also be written in
trace form, as
\beq
\label{eq:cs-finite}
\mcs_a=-\frac{e^2}{3\hbar cV}\e_a\epsilon_{ijk}\Tri\,[Pr_iPr_jPr_k].
\eeq
\endnumparts
We first recast the orbital magnetization \eref{eq:M-finite} as
\beq
M_a=-\frac{e}{2cV}\epsilon_{abc}\mathrm{Tr}\,[Pr_bv_c]
=\frac{e}{2\hbar cV}\epsilon_{abc}\Tri\,\bigl[Pr_b\ch^0r_c\bigr]
\eeq
and then subtract \eref{eq:M-lcfin} and \eref{eq:M-icfin}
from it to find, after some manipulations,  
\beq
M^{\mathrm{CS}}_a=
-\frac{e}{\hbar cV}\epsilon_{abc}\mathrm{Im\,Tr}\,[Q\ch^0Pr_bPr_c].
\eeq
Replacing $\ch^0$ with $\ch-e\e_dr_d$ and using $Q\ch P=0$, 
\beq \mcs_a
=-\frac{e^2}{\hbar cV}\epsilon_{abc}\e_d\mathrm{Im\,Tr}\,[Pr_dPr_bPr_c].  
\eeq 
The imaginary part of the trace vanishes if any two of the indices
$b$, $c$, or $d$ are the same, and therefore $d$ must be equal to $a$.
Using the cyclic property 
we conclude that all non-vanishing terms in the sum over $b$ and $c$
are identical, leading to \eref{eq:cs-finite}.  This part of the
field-induced magnetization is clearly isotropic, with a coupling
strength (see equation~\eref{eq:alpha-iso}) given by
\beq
\label{eq:theta-cs-finite}
\tcs=-\frac{4\pi^2}{3V}\epsilon_{ijk}\Tri[Pr_iPr_jPr_k].
\eeq
This expression can assume nonzero
values because the Cartesian components of the projected position
operator $P\r P$ do not commute~\cite{mv97}.

We have used \eref{eq:M-lcfin}--\eref{eq:cs-finite} to evaluate
the OMP contributions $\alc$, $\aic$, and $\acs$ for finite samples,
finding excellent agreement with the $k$-space calculations using
\eref{eq:M-lc}--\eref{eq:M-cs}. As an example, the finite-sample results 
for the $zz$ component are plotted as the symbols in figure~\ref{fig:al_zz_c}.

\section{Linear-response expression for the OMP tensor}
\label{sec:linear-response}

In sections~\ref{sec:finite-field-k} and \ref{sec:finite-field-k-inv}
expressions were given for evaluating $\mb(\eb)$ under periodic
boundary conditions.  Used in conjunction with finite-field
{\it ab-initio} methods for periodic insulators~\cite{umari02,souza02},
they allow to calculate the OMP tensor by finite differences.  
Alternatively, the electric field may be treated
perturbatively~\cite{nunes01}. With this approach in mind, we shall now
take the $\eb$-field derivative in \eref{eq:alpha} analytically
and obtain an expression for the OMP tensor which is amenable to
density-functional perturbation-theory implementation~\cite{baroni01}.
It should be kept in mind that in the context of self-consistent-field
(SCF) calculations the ``zero-field'' part of the Hamiltonian
(\ref{eq:ham}),
\beq
\label{eq:h-scf}
\ch^0=-\frac{\hbar^2}{2m}\nabla^2+V_{\rm SCF}(\r),
\eeq 
does depend on $\eb$ implicitly, through the charge density.  As we
will see, this dependence gives rise to additional local-field
screening terms in the expression for the OMP.

\begin{figure}
\centering\includegraphics{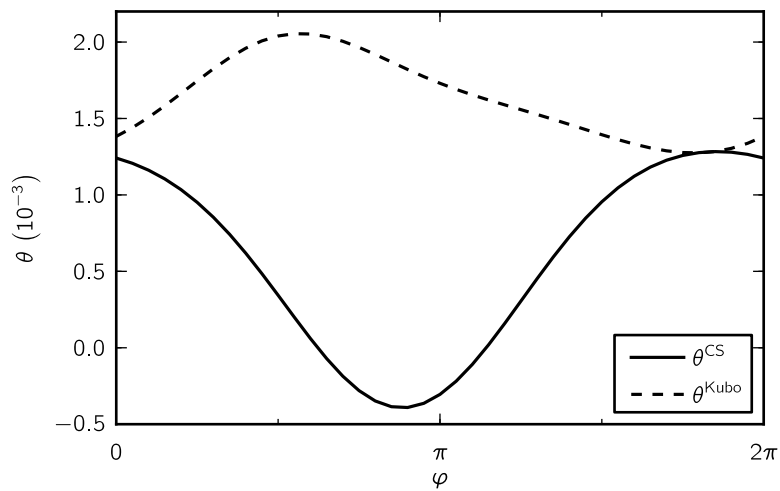}
\caption{Contributions to the isotropic OMP from the
  Chern-Simons term $\acs$ and from the Kubo-like terms $\alc$ and
  $\aic$, expressed in terms of the dimensionless coupling strength
  $\theta$ in \eref{eq:alpha-iso}. Model parameters are the same
  as for figure~\ref{fig:al_zz}.}
\label{fig:theta}
\end{figure}

We shall only consider the case where the OMP is calculated for a
reference state at zero field, which we indicate by a superscript
``0.'' Upon inserting \eref{eq:M-tot} into \eref{eq:alpha} we obtain
the three gauge-invariant OMP terms in \eref{eq:omp-tot}.  The term
$\acs$ is clearly of the isotropic form \eref{eq:alpha-iso}, with
\numparts
\beq
\label{eq:theta-cs}
\tcs=-\frac{1}{4\pi}\int \rmd^3k\,\epsilon_{ijk}\tr\left[A_i^0\partial_jA_k^0-
\frac{2\rmi}{3}A_i^0A_j^0A_k^0\right].
\eeq
This is the same expression as obtained previously by
heuristic methods~\cite{qi08,essin09}\footnote{An inconsistency
in the published literature regarding the numerical prefactor in
\eref{eq:theta-cs} has been resolved: see~\cite{essin09E}}.
The other two terms were not considered in the previous works. They are
\begin{eqnarray}
\label{eq:alpha-lc}
\alc_{da}=\gamma\epsilon_{abc}\int [\rmd k]\,
\sum_n^N\,\mathrm{Im}
\Big(
  2\bra{\partb u^0_{n\k}}(\partial_c\hk)\ket{\wt{\partial}_Du_{n\k}^0}\nonumber\\
\;\;\;\;\;\;\;\;\;\;\;\;\;\;\;\;\;\;\;\;\;\;\;\;\;\;\;\;\;\;\;\;\;\;\;\;\;\;\;
\;\;
-\bra{\partb u^0_{n\k}}(\partial_D \hk)\ket{\wt{\partial}_cu_{n\k}^0}
\end{eqnarray}
and
\begin{eqnarray}
\label{eq:alpha-ic}
\aic_{da}=\gamma\epsilon_{abc}\int [\rmd k]\,
\sum_{mn}^N\,\mathrm{Im}
\Big(
  2\bra{\partb u_{n\k}^{0}}\wt{\partial}_D u_{m\k}^0\rangle 
  \bra{u_{m\k}^0}(\partial_c\hk)\ket{u_{n\k}^0}\nonumber\\
\;\;\;\;\;\;\;\;\;\;\;\;\;\;\;\;\;\;\;\;\;\;\;\;\;\;\;\;\;\;\;\;\;\;\;\;\;\;\;\,\,
-\bra{\widetilde{\partial}_b u_n^0}\widetilde{\partial}_c u_m^0\rangle
      \bra{u_m^0}(\partial_D \hk)\ket{u_n^0}
\Big),
\end{eqnarray}
\endnumparts
where $\partial_D$ denotes the field-derivative $\partial/\partial
\e_d$ and
\beq
\label{eq:proj-field-der}
\ket{\wt{\partial}_D u_{n\k}^0}\equiv
Q_\k\left. \ket{\partial_D u_{n\k}}\right|_{\eb=0}
\eeq
are the first-order field-polarized states projected onto the
unoccupied manifold. The terms containing $\partial_D \hk$ describe
the screening by local fields. They vanish for tight-binding models
such as the one in this work, but should be included in
self-consistent calculations, in the way described in \cite{baroni01}.
We shall sometimes refer to $\alc$ and $\aic$ as `Kubo' contributions
because, unlike the Chern-Simons term, they involve first-order
changes in the occupied orbitals and Hamiltonian, in a manner
reminiscent of conventional linear-response theory.\footnote{The
  terminology `Kubo terms' for $\alc$ and $\aic$ is only meant to be
  suggestive. A Kubo-type linear-response calculation of the OMP
  should produce all three terms, including $\acs$.}

Equations~\eref{eq:theta-cs}--\eref{eq:alpha-ic} are the main result
of this section.  The derivation of \eref{eq:alpha-lc} and
\eref{eq:alpha-ic} is somewhat laborious and is sketched in
\ref{app:derivation}.  We emphasize that the Kubo-like terms, besides
endowing the tensor $\alpha$ with off-diagonal elements, also
generally contribute to its trace, which therefore is not purely
geometric. Writing the isotropic part of the OMP response in the form
\eref{eq:alpha-iso}, we then have
\beq
\theta=\tcs+\tkubo.
\eeq
The two contributions are plotted for our model in
figure~\ref{fig:theta}.  Moreover, the open circles in
figure~\ref{fig:al_zz} show the $zz$ and $zy$ components of the OMP
tensor computed from \eref{eq:theta-cs}--\eref{eq:alpha-ic},
confirming that the analytic field derivative of the magnetization was
taken correctly.

In the case of an insulator with a single valence band, the partition
\eref{eq:omp-tot} of the OMP tensor acquires some interesting
features. The terms $\aic$ and $\acs$ become purely itinerant, i.e.,
they vanish in the limit of a crystal composed of non-overlapping
molecular units, with one electron per molecule.  Also, the first term
in the expression \eref{eq:alpha-ic} for $\aic$ --~the only term for
tight-binding models~-- becomes traceless, as can be readily verified
in a Hamiltonian gauge (where
$\hk\ket{u^0_{n\k}}=E^0_{n\k}\ket{u^0_{n\k}}$) with the help of the
perturbation theory formula~\cite{nunes01}
\beq
\label{eq:sternheimer-efield}
\ket{\wt{\partial}_Du_{n\k}^0}=\rmi e\sum_{m>N}
\frac{\ket{u_{m\k}^0}\bra{u_{m\k}^0}}{E_n^0-E_m^0}\ket{\partial_d u_{n\k}^0}.
\eeq
In order to verify these features numerically, we calculated the
various contributions treating only the lowest band of our
tight-binding model as occupied. The molecular limit was taken by
setting to zero the hoppings between neighbouring eight-site cubic
``molecules.''

\begin{figure}
\centering\includegraphics{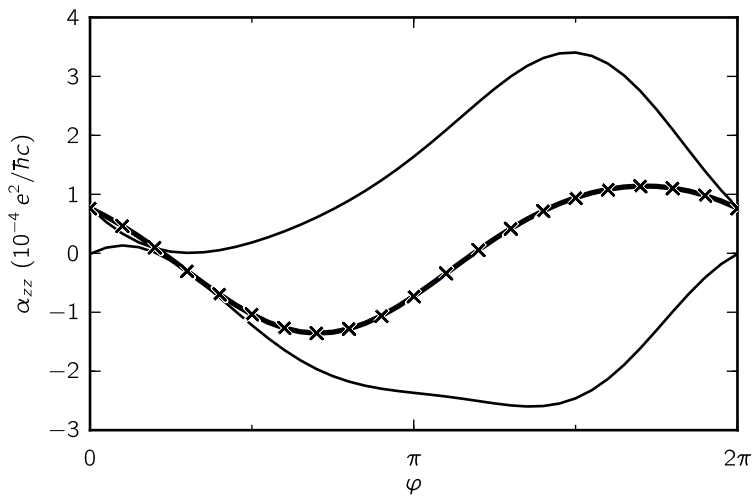}
\caption{Comparison between $\alpha_{zz}$ calculated treating the two
  lowest bands as occupied (crosses) and the sum
  $\alpha_{zz}^{(1)}+\alpha_{zz}^{(2)}$ (thick solid line), where
  $\alpha_{zz}^{(n)}$ (thin solid lines) correspond to treating only
  the lowest band ($n=1$, upper line) or the second-lowest band
  ($n=2$, lower line) as occupied. Model parameters are the same as
  for figure~\ref{fig:al_zz} except that the second lowest on-site energy 
  in table~\ref{tab:tb_model} is raised from $-$6.0 to $-$5.0 
  in order to keep the two lowest bands well-separated.}
\label{fig:bsc}
\end{figure}

It could have been anticipated from the outset that
the Chern-Simons term \eref{eq:theta-cs} could not be the
entire expression for the OMP, based on the following
argument~\cite{essin10}.
Consider an insulator with $N>1$ valence bands, all of which are
isolated from one another. By looking at $\alpha_{da}$ as
$\partial P_d/\partial B_a$ one can argue that, since
each band carries a certain amount of polarization ${\bi P}^{(n)}$,
the total OMP should satisfy the relation
\beq
\label{eq:bsc}
\alpha=\sum_n^N\,\alpha^{(n)},
\eeq
where $\alpha^{(n)}$ is the OMP one would obtain by filling band $n$
while keeping all other bands empty. We shall refer to this property
as the ``band-sum-consistency'' of the OMP. It only holds exactly for
models without charge self-consistency (see the analytic proof in
\ref{app:band-sum-consistency}), but that suffices for the purpose of
the argument.  We note that the Chern-Simons contribution
\eref{eq:theta-cs} alone is {\it not} band-sum-consistent, because the
second term therein vanishes for a single occupied band. Hence an
additional contribution, also band-sum-inconsistent, must necessarily
exist. Indeed, both $\alc$ and $\aic$ are band-sum-inconsistent, in
such a way that the total OMP satisfies \eref{eq:bsc}.  This is
illustrated in figure~\ref{fig:bsc} for our tight-binding model.
%

\section{Summary and outlook}

In summary, we have developed a theoretical framework for calculating
the frozen-ion orbital-magnetization response (OMP) to a static
electric field. This development fills an important gap in the
microscopic theory of the magnetoelectric effect, paving the way to
first-principles calculations of the full response.  While the OMP is
often assumed to be small compared to the lattice-mediated and
spin-magnetization parts of the ME response, there is no {\it a
  priori} reason why it should always be so. In fact, in strong $Z_2$
topological insulators it is the only contribution that survives, and
the predicted value is large compared to that of prototypical
magnetoelectrics. Although the measurement of the $\theta=\pi$ ME
effect in topological insulators is challenging, as time-reversal
symmetry must be broken to gap the
surfaces~\cite{qi08,essin09,hasan10}, there may be other related
materials where those symmetries are broken already in the bulk.  The
present formalism should be helpful in the ongoing computational
search for such materials with a large and robust OMP.

A key result of this work is a $k$-space expression for the orbital
magnetization of a periodic insulator under a finite electric field
$\eb$ (equations \eref{eq:M-tot-a}--\eref{eq:M-ice-bulk}, or
equivalently, \eref{eq:M-tot}--\eref{eq:M-cs}).  In addition to the
terms \eref{eq:M-lc}--\eref{eq:M-ic} already present at zero
field~\cite{ceresoli06}, in three dimensions the field-dependent
magnetization comprises an additional purely isotropic `Chern-Simons'
term, given by \eref{eq:M-cs}. This new term depends explicitly on
$\eb$ and only implicitly on $\hk$, while the converse is true for the
other terms.  Moreover, it is a multivalued quantity, with a quantum
of arbitrariness $\mb_0=\eb e^2/hc$ along $\eb$.  Thus, the analogy
with the Berry-phase theory of electric
polarization~\cite{King-Smith,resta-review07}, where a similar
  quantum arises, becomes even more profound at finite electric
field.

The Chern-Simons term $\mcsb$ is responsible for the geometric part of
the OMP response discussed in \cite{qi08,essin09} in connection with
topological insulators.  We have clarified that in materials with
broken time-reversal and inversion symmetries in the bulk the CSOMP
does not generally constitute the full response, as the remaining
orbital magnetization terms, $\lctb$ and $\ictb$, can also depend
linearly on $\eb$. Their contribution to the OMP, given by
\eref{eq:alpha-lc}--\eref{eq:alpha-ic}, is twofold: (i) to modify the
isotropic coupling strength $\theta$; and (ii) to introduce an
anisotropic component of the response.

Another noteworthy result is equation~\eref{eq:theta-cs-finite} for
the Chern-Simons OMP of finite systems.  One appealing feature of this
expression is that it allows one to calculate the CSOMP without the
need to choose a particular gauge. Instead, its $k$-space counterpart,
equation~\eref{eq:theta-cs}, requires for its numerical evaluation a
smoothly varying gauge for the Bloch states across the Brillouin zone.
\Eref{eq:theta-cs-finite} is also the more general of the two, as it
can be applied to noncrystalline or otherwise disordered systems.

We conclude by enumerating a few questions that are raised by the
present work. 
Do the individual gauge-invariant OMP terms identified here in a
  one-electron picture remain meaningful for interacting systems, and
  can they be separated experimentally?  (This appears to be the case
  for $\lctb$ and $\ictb$ at $\eb=0$~\cite{souza08}.)  How does the
expression \eref{eq:theta-cs}--\eref{eq:alpha-ic} for the linear OMP
response change when the reference state is under a finite electric
field $\eb$?  Finally, we note that equation~\eref{eq:cs-finite} for
the CSOMP of finite systems has a striking resemblance to a formula
given by Kitaev~\cite{kitaev06} for the 2-D Chern invariant
characterizing the integer quantum Hall effect. Can this connection
be made more precise, in view of the fact that the quantum of
indeterminacy in $\tcs$ is associated with the possibility of changing
the Chern invariant of the surface layers?  These questions are left
for future studies.

\ack
This work was supported by NSF Grant Nos. DMR 0706493 and 0549198.
Computational resources have been provided by NERSC. The authors
gratefully acknowledge illuminating discussions with Andrew Essin,
Joel Moore, and Ari Turner.

\appendix

\section{Tight-binding model and technical details}
\label{app:model}

\subsection*{Tight-binding model}

We have chosen for our tests a model of an ordinary (that is,
non-topological) insulator. The prerequisites were the following.  It
should break both time-reversal and inversion symmetries, as the
OMP tensor otherwise vanishes identically.  It should be three-dimensional,
as the geometric part of the response vanishes otherwise. Its symmetry
should be sufficiently low to render all nine components of the OMP
tensor nonzero.  Finally, it should have multiple valence bands, for
generality.

\begin{figure}
\centering\includegraphics{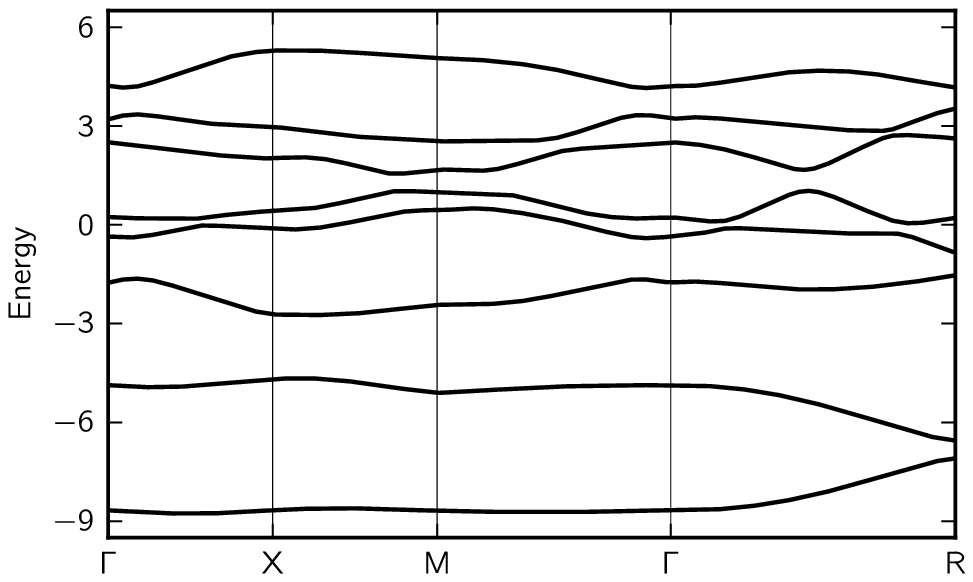}
\caption{ Band structure of the cubic-lattice tight-binding model
  given by \eref{eq:ham-tb}, for the choice of parameters in
  table~\ref{tab:tb_model} and $\varphi=0$.
}
\label{fig:bands}
\end{figure}

We opted for a spinless model on a cubic lattice. It can be obtained
starting with a one-site simple cubic model, doubling the cell in each
direction, and assigning random on-site energies $E_{\i}$ and complex
first-neighbour hoppings $t_{\j\to\i}=t\rme^{\rmi\phi_{\j\to\i}}$ of fixed
magnitude $t=1$. The Hamiltonian reads
\beq
\label{eq:ham-tb}
\ch^0=\sum_{\i}\,E_{\i}c^{\dagger}_{\i}c^{\ph{\dagger}}_{\i}+
\sum_{\langle\i\j\rangle}\,\rme^{\rmi\phi_{\j\to\i}}
c^{\dagger}_{\i}c^{\ph{\dagger}}_{\j},
\eeq
where ${\bi i}=(x,y,z)$ labels the sites and $\langle {\bi i\bi j}\rangle$
denotes pairs of nearest-neighbour sites.  The values of $E_{\i}$ on two
of the eight sites were adjusted to ensure a finite gap everywhere in
the Brillouin zone between the two lowest bands (chosen as the valence
bands) and the remaining six. We also made sure that nonzero phases
$\phi_{\j\to\i}$ were not restricted to two-dimensional square-lattice
planes, otherwise those are mirror symmetry planes, whose existence is
sufficient to make the diagonal elements of the OMP tensor vanish.  In
our calculations all the model parameters were kept fixed except for one
phase, which was scanned over the range $[0,2\pi]$, and the results
are plotted as a function of this phase $\varphi$.  For reference, the
on-site energies and the phases of the hopping amplitudes are listed
in table~\ref{tab:tb_model}. The energy bands are shown in
figure~\ref{fig:bands} for $\varphi=0$.

In order to couple the system to the electric field and to be able to
define its orbital magnetization, the position operator $\r$ must be
specified along with $\ch^0$.  We have chosen the
simplest representation where $\r$ is diagonal in the tight-binding
basis.

\begin{table}
  \caption{\label{tab:tb_model}
    The parameters of the tight-binding model. Columns I--III give the site 
    coordinates $\i=(x,y,z)$, in units of the lattice constant $a=1$ of
    the $2\times 2\times 2$ primitive cubic cell. Column IV contains the on-site 
    energies $E_{\i}$, and the last three columns contain the phases of 
    the complex nearest-neighbour hopping amplitudes along bonds in the
    negative $\hat{\bi x}$, $\hat{\bi y}$, and $\hat{\bi z}$ directions.
  }
\begin{indented}
\item[]\begin{tabular}{@{}rrrlccc}
\br
$x$ & $y$ & $z$ & \ph{$-$}$E_{\i}$ & 
$\phi_{\mbox{\scriptsize $(\i+\hat{\bi x}/2)\to\i$}}$ & 
$\phi_{\mbox{\scriptsize $(\i+\hat{\bi y}/2)\to\i$}}$ & 
$\phi_{\mbox{\scriptsize $(\i+\hat{\bi z}/2)\to\i$}}$ \\ 
\mr
0.0 & 0.0 & 0.0 & $-$6.5 & $\varphi\in[0,2\pi]$ & 0.5$\pi$ & 1.7$\pi$ \\
0.5 & 0.0 & 0.0 & \ph{$-$}0.9 & 1.3$\pi$ & 0.2$\pi$ & 0.5$\pi$ \\
0.5 & 0.5 & 0.0 & \ph{$-$}1.4 & 0.8$\pi$ & 1.4$\pi$ & 0.6$\pi$ \\
0.0 & 0.5 & 0.0 & \ph{$-$}1.2 & 0.3$\pi$ & 1.9$\pi$ & 1.0$\pi$ \\
0.0 & 0.0 & 0.5 & $-$6.0\footnotemark[1] & 1.4$\pi$ & 0.8$\pi$ & 0.3$\pi$ \\
0.5 & 0.0 & 0.5 & \ph{$-$}1.5 & 0.6$\pi$ & 1.7$\pi$ & 0.7$\pi$ \\
0.5 & 0.5 & 0.5 & \ph{$-$}0.8 & 0.8$\pi$ & 0.6$\pi$ & 1.2$\pi$ \\
0.0 & 0.5 & 0.5 & \ph{$-$}1.2 & 1.9$\pi$ & 0.3$\pi$ & 1.4$\pi$ \\
\br
\end{tabular}
\item[]\footnotemark[1] In figure~\ref{fig:bsc} the value $-$5.0 was used instead.
\end{indented}
\end{table}

\subsection*{Technical details}

The calculations employing periodic boundary conditions were carried
out on an $80\times80\times80$ $k$-point mesh, and the $k$-space
implementation of finite electric fields was done using the method
discussed in section~V of \cite{souza04}.  The open boundary
condition calculations used cubic samples containing $L\times L\times
L$ eight-site unit cells, that is, $2L+1$ sites along each edge.  For
large $L$, we expect the magnetization to scale as
\beq
\label{eq:fit} 
\mb(L)=\mb+\frac{\ab}{L}+\frac{\bb}{L^2}+\frac{\cb}{L^3}, 
\eeq 
where $\ab$, $\bb$,
and $\cb$ account for face, edge, and corner corrections,
respectively~\cite{ceresoli06}. Calculations of $\mb(L)$ under small
fields were done using $L=4,5,6,7$, 
and then fitted to \eref{eq:fit} in order to extract the value
$\mb$ of the magnetization in the $L\to\infty$ limit.
The differences between OMP values calculated in various ways as shown
in figures~\ref{fig:al_zz}~and~\ref{fig:al_zz_c} were of the order
of $10^{-7}$\,$e^2/\hbar c$ or less.

Before evaluating the $k$-space expressions for $\mb(\eb)$
[\eref{eq:M-tot-a}--\eref{eq:M-ice-bulk} and
\eref{eq:M-lc}--\eref{eq:M-cs}] and $\alpha$
[\eref{eq:theta-cs}--\eref{eq:alpha-ic}] on a grid, they need to be
properly discretized. The presence of the gauge-dependent Berry
connection in \eref{eq:M-ice-bulk} demands the use of a ``smooth
gauge'' for its evaluation, where the valence Bloch states given by
\eref{eq:gauge} are smoothly varying functions of $\k$.  This is
achieved by projecting a set of trial orbitals onto the set of
occupied Bloch eigenstates according to the prescription in
equations~(62--64) of \cite{mv97}. (For the tight-binding model
discussed below, when treating the two lowest bands as occupied, the
two trial orbitals are chosen as delta functions located at the two
sites with lowest on-site energy.) If needed, this one-shot projection
procedure can be improved upon by finding an optimally smooth gauge
using methods based on minimizing the real-space spread of the
WFs~\cite{mv97}, but we found our results to change negligibly when
performing this extra step.  In a smooth gauge the needed
$k$-derivatives of the Bloch states and of the Berry connection matrix
are then evaluated by straightforward numerical differentiation. Note
that \eref{eq:M-lc-bulk}~and~\eref{eq:M-ic0-bulk} should be evaluated
in the same smooth gauge as \eref{eq:M-ice-bulk}, as these three
equations are not separately gauge-invariant.  A smooth gauge must
also be used for \eref{eq:M-cs}~and~\eref{eq:theta-cs}, because, as
discussed in section~\ref{sec:finite-field-k-inv}, the Chern-Simons
3-form is locally gauge-dependent.

The same strategy can be used to discretize \eref{eq:M-lc} and
\eref{eq:M-ic}. However, since the $k$-derivatives appearing in those
equations are covariant, the discretized form of the covariant
derivative \eref{eq:cov-der} given in \cite{souza04,ceresoli06} may be
used instead, circumventing the need to work in a smooth gauge. We
have implemented both approaches, finding excellent agreement between
them.

Finally we come to equations~\eref{eq:alpha-lc}~and~\eref{eq:alpha-ic}.
In addition to the $k$-derivative of the valence Bloch states, we need
their (covariant) field-derivative \eref{eq:proj-field-der}, as well
as the $k$-derivative of $\hk$. The latter quantity is easily
calculated within the tight-binding method, and for the former we used
the linear-response expression \eref{eq:sternheimer-efield}.  Note
that this requires choosing the
unperturbed states to be in
the Hamiltonian gauge.  This choice precludes calculating the
$k$-derivative on the right-hand-side of
\eref{eq:sternheimer-efield} by straightforward finite
differences, which can only be done in a smooth gauge. But because
$\ip{u^0_{m\k}}{\partial_du^0_{n\k}}$ equals
$\ip{u^0_{m\k}}{\wt{\partial}_du^0_{n\k}}$ for $m>N$, the discretized
covariant derivative approach may be used instead.  Alternatively, one
can evaluate the ordinary $k$-derivative by summation over states as
\beq
\label{eq:sum-states}
\ket{\partial_d u_{n\k}^0}=\sum_{m\not= n}\,
\ket{u_{m\k}^0}
\frac{\bra{u_{m\k}^0}(\partial_d \hk)\ket{u_{n\k}^0}}{E_{n\k}^0-E_{m\k}^0}.
\eeq
We note that this formula may not be used to calculate the geometric
term \eref{eq:theta-cs}, because it induces locally a parallel
transport gauge ($A_{nn}^0=0$), which cannot be enforced globally
since the Brillouin zone is a closed space.

\section{Derivation of equations \eref{eq:alpha-lc} and
  \eref{eq:alpha-ic}}
\label{app:derivation}

For notational simplicity we drop the crystal momentum index $\k$. So,
for example, $\ket{u_{n\k}}$ shall be denoted by $\ket{u_n}$.  In
order to calculate the OMP terms
\beq
\label{eq:alc-def}
\alc_{da}=
\left.
  \partial_D\wt{M}_a^{({\rm LC})}
  \right|_{\eb=0}
\eeq
 and 
\beq
\label{eq:aic-def}
\aic_{da}=
\left.
  \partial_D\wt{M}_a^{({\rm IC})}
  \right|_{\eb=0}
\eeq
starting from \eref{eq:M-lc-a} and \eref{eq:M-ic-a}, we shall
first examine the field- and $k$-derivatives of certain basic
quantities.  

We begin by noting that the field-derivative $\partial_D P=-\partial_D
Q$ of the projection operator \eref{eq:proj} can be written as
\beq
\label{eq:dP}
\partial_DP
=\sum_n^N
\left(
\ket{\wt{\partial}_Du_n}\bra{u_n}+\ket{u_n}\bra{\wt{\partial}_Du_n}
\right)
\equiv\wt{\partial}_DP,
\eeq
in terms of the covariant field-derivative \eref{eq:proj-field-der}
(a similar expression holds for the $k$-derivative).
This follows from a relation analogous to \eref{eq:cov_deriv}:
\beq
\label{eq:fullderiv}
\ket{\partial_Du_n}=
\ket{\wt{\partial}_Du_n}-\rmi\sum_l^NA_{ln,D}\ket{u_l},
\eeq
where
\beq
\label{eq:A-fld}
A_{ln,D}=\rmi\bra{u_l}\partial_D u_n\rangle=A_{nl,D}^*
\eeq
is the Berry connection matrix along the parametric direction $\e_d$.
With the help of \eref{eq:dP} the field-derivative of 
\eref{eq:metric-curv} becomes
\beq
\label{eq:dFnmbc}
\fl\partial_D F_{nm,bc}=\bra{\partial^2_{Db} u_n}Q\ket{\partial_c u_m}+
          \bra{\partial_b u_n}Q\ket{\partial^2_{Dc} u_m}\nonumber 
          + \rmi(F_{bD}A_c)_{nm}-\rmi(A_bF_{Dc})_{nm},
\eeq
where $F_{bD}$ is obtained from $F_{bd}$ by replacing
$\partial_d$ with $\partial_D$.  We shall also need the
field- and $k$-derivatives of the matrix $H^0_{nm}$ defined by \eref{eq:h0}:
\beq
\label{eq:dHnm}
\left.\partial_D(H_{nm}^0)\right|_{\eb=0}=\rmi\left[A_D^0,\h0\right]_{nm}+
\left.(\partial_DH^0_{\mathrm{op}})_{nm}\right|_{\eb=0}
\eeq
\beq
\label{eq:dcHnm}
\left.\partial_c(H_{nm}^0)\right|_{\eb=0}=
\rmi\left[A_c^0,\h0\right]_{nm}+
\left.(\partial_cH^0_{\mathrm{op}})_{nm}\right|_{\eb=0},
\eeq
where we introduced the notation
$(\partial_{D,c}H^0_{\mathrm{op}})_{nm}\equiv\bra{u_n}\partial_{D,c}H^0\ket{u_m}$,
where `op' indicates that the derivative is taken on the operator
itself, not its matrix representation.  These two relations follow
directly from \eref{eq:fullderiv} and \eref{eq:cov_deriv}.  We will
also make use of identities such as
\beq
\mathrm{Re\,tr}\left[XF_{bc}\right]=
\mathrm{Re\,tr}\left[X^{\dagger}F_{cb}\right].
\eeq
In particular, if $X$ and $Y$ are Hermitian,
\beq
\label{eq:Retr}
\mathrm{Re\,tr}\left[XYF_{bc}\right]=
\mathrm{Re\,tr}\left[YXF_{cb}\right].
\eeq

We are now ready to evaluate \eref{eq:aic-def}: 
\beq
\label{eq:OMh_1}
\fl\aic_{da}=-\gamma\epsilon_{abc}\int
[\rmd k]\tri
\bigl[F_{bc}\partial_D\h0 
+\left.\h0\partial_DF_{bc}\bigr]\right|_{\eb=0}.  
\eeq
Inserting \eref{eq:dFnmbc} and \eref{eq:dHnm} on the right-hand
side generates a number of terms.  Some can be combined upon
interchanging dummy indices $b\leftrightarrow c$
and invoking \eref{eq:Retr}, leading to
\begin{eqnarray} 
\fl\aic_{da}=-\gamma\epsilon_{abc}\int[\rmd k]
\Big(
  2\mathrm{Re\,tr} \left[A_D\h0F_{bc}+\h0F_{bD}A_c\right] 
  +\tri \left[ F_{bc}\partial_D \h0_{\rm op}\right]
\nonumber \\
\;\;\;\;\;\;\;\;  
  +2\mathrm{Im}\sum_{mn}^N\h0_{mn}\bra{\partial^2_{Db}u_n}
Q\ket{\partial_cu_m}\Big)\Big|_{\eb=0}.  
\end{eqnarray}
Integrating the last term by parts in $k_b$ and using
\eref{eq:dP} and \eref{eq:dcHnm} again produces a number of
terms, most of which cancel out. The end result reads
\beq
\label{eq:alpha-ic-alt}
\aic_{da}= \gamma\epsilon_{abc}\int [\rmd k]\,\tri
\left.\left[
            2F_{bD}\partial_cH^0_{\mathrm{op}}
           -F_{bc}\partial_D\h0_{\rm op}
      \right]\right|_{\eb=0}.  \eeq
Similarly, \eref{eq:alc-def} can be evaluated by repeatedly using
\eref{eq:dP} and integrating by parts the terms with mixed field- and
$k$-derivatives, yielding
\beq
\label{eq:alpha-lc-alt}
\alc_{da}=
\gamma\epsilon_{abc}\int [\rmd k]\,\tri
\left.\left[
            2(\partial_c\h0)_{bD}-(\partial_D\h0)_{bc}
      \right]\right|_{\eb=0},
\eeq
where $(\partial_cH^0)_{bD}$ and $(\partial_D\h0)_{bc}$
are defined in analogy with \eref{eq:Xab}, e.g.,
\beq
(\partial_cH^0)_{nm\k,bD}=
\me{\partb u_{n\k}}{(\partial_c\hk)}{\wt{\partial}_D u_{m\k}}.
\eeq
Equations~\eref{eq:alpha-lc-alt}~and~\eref{eq:alpha-ic-alt} are
respectively equivalent to \eref{eq:alpha-lc} and
\eref{eq:alpha-ic} in the main text. The gauge invariance of these
equations follows from the fact that they are written as traces over
gauge-covariant objects. (We also note that the covariant derivative
transforms according to \eref{eq:gauge} regardless of the
parameter with respect to which the differentiation is carried out.)

\section{Band-sum consistency of the OMP}
\label{app:band-sum-consistency}

Here we show analytically that the OMP tensor $\alpha$ satisfies the
band-additivity relation \eref{eq:bsc} in models without charge
self-consistency. In order to isolate the contribution $\alpha^{(n)}$
coming from valence band $n$ (assumed to be well-separated in energy
from all other bands), we choose the Hamiltonian matrix to be diagonal
at zero field, i.e., $H^0_{mn\k}(\eb=0)=E_{n\k}^0\delta_{mn}$.  If in
addition we use a parallel-transport gauge for the linear electric
field perturbation~\cite{nunes01} (this is achieved by setting to zero
the matrix $A_D$ defined in \eref{eq:A-fld}) we find, using
\eref{eq:dHnm}, $\partial_D H^0_{mn\k}|_{\eb=0}=0$.  With the help of
these two relations, the field-derivative $\partial_D M_a|_{\eb =0}$
of \eref{eq:M-tot-a} is easily taken.  From the first two terms
therein we obtain (dropping the index $\k$)
\beq
\left.2\gamma\epsilon_{abc}\int [\rmd k]\,\sum_n^N\,
\im\bra{\partial_b u_n}\partial_c(\h0+E_n)\ket{\partial_D u_n}
\right|_{\eb=0}.
\eeq
In the parallel-transport gauge $\ket{\partial_D u_n}$ is given by
\eref{eq:sternheimer-efield}, and combining the resulting
expression with the field-derivative of the third term in
\eref{eq:M-tot-a} yields
\begin{eqnarray}
\label{eq:omp-h-gauge}
\fl \alpha_{da}= 2e\gamma\epsilon_{abc}\int [\rmd k]\,\sum_n^N\,
  \re\bra{\partial_b u_n^0}\partial_c(\h0+E_n^0) 
  \left(
    \sum_{l>N}\,\frac{\ket{u_l^0}\bra{u_l^0}}{E_n^0-E_l^0} \right)
  \ket{\partial_d u_n^0}\nn
  -e\gamma\epsilon_{abc}\,\int [\rmd k]\,\sum_{mn}^N\,
  \re\left(\bra{u_m^0}\partial_d u_n^0\rangle\bra{\partial_b
      u_n^0}\partial_c u_m^0\rangle\right).
\end{eqnarray}
To find $\alpha_{da}^{(n)}$ we replace $\sum_{l>N}$ with $\sum_{l\not=
  n}$ and reduce sums $\sum_{mn}^N$ and $\sum_n^N$ to single terms.
Inserting these expressions into \eref{eq:bsc} and
splitting $\sum_{l\not= n}$ into $\sum_{l>N}$ and $\sum_{l\not=
  n}^N$, some terms cancel and others can be combined, leading to
\begin{eqnarray}
\label{eq:conjecture-b}
\fl \epsilon_{abc}\int [\rmd k]\,\sum_n^N\,\sum_{m\not= n}^N\,\re
\Biggl[
  \bra{u_m^0}\partial_d u_n^0\rangle \nn
  \times\left(
    \frac{\bra{\partial_b u_n^0}\partial_c(\h0+E_n^0)\ket{u_m^0}}{E_n^0-E_m^0}+
    \frac{1}{2}\bra{\partial_b u_n^0}\partial_c u_m^0\rangle
  \right)
\Biggr]=0.
\end{eqnarray}
The LHS is proportional to the difference between $\alpha_{da}$ and
$\sum_n^N\,\alpha_{da}^{(n)}$, and vanishes as a result of an exact
cancellation between the terms $(n,m)$ and $(m,n)$ in the double
sum. The integrand of the $(n,m)$ term is
\beq
\label{eq:nm-term}
\epsilon_{abc}\re
\left[
  \bra{u_m^0}\partial_d u_n^0\rangle
  \left(
    \frac{\bra{\partial_b u_n^0}\partial_c(\h0+E_n^0)\ket{u_m^0}}{E_n^0-E_m^0}
    + \frac{1}{2}\bra{\partial_b u_n^0}\partial_c u_m^0\rangle
  \right)
\right],
\eeq
and after some manipulations the integrand of the $(m,n)$ term becomes
\beq
\label{eq:mn-term}
\epsilon_{abc}\re
\left[
  \bra{u_m^0} \partial_du_n^0\rangle
  \left(
    \frac{\bra{u_n^0}\partial_c(\h0+E_m^0)\ket{\partial_b u_m^0}}{E_n^0-E_m^0}
    +\frac{1}{2}\bra{\partial_b u_n^0}\partial_c u_m^0\rangle
  \right)
\right].
\eeq
The final step is to use the identity
\beq
\fl\frac{\bra{\partial_b u_n^0}\partial_c(\h0+E_n^0)\ket{u_m^0}}{E_n^0-E_m^0}
=\frac{\bra{\partial_b u_n^0}E_m^0-\h0\ket{\partial_c u_m^0}}{E_n^0-E_m^0}
-\partial_c(E_n^0+E_m^0)\frac{\bra{u_n^0}\partial_b
  u_m^0\rangle}{E_n^0-E_m^0}.
\eeq 
(This identity follows from the relation 
\beq
\label{eq:sternheimer-partial-k}
(\h0-E_m^0)\ket{\partial_c u_m^0}=
-(\partial_c \h0-\partial_c E_m^0)\ket{u_m^0},
\eeq
which in turn can be obtained by expanding
$\h0\ket{u_m^0}=E_m^0\ket{u_m^0}$ to first order in the change in
wavevector $\k$.)  The quantity \eref{eq:nm-term}+\eref{eq:mn-term}
then becomes
\beq
\fl\epsilon_{abc}\re
\left[
  \bra{u_m^0}\partial_d u_n^0\rangle
  \left(
  \frac{\bra{\partial_b u_n^0}E_m^0-\h0\ket{\partial_c u_m^0}}{E_n^0-E_m^0}
  +\frac{\bra{\partial_c u_n^0}E_n^0-\h0\ket{\partial_b u_m^0}}{E_n^0-E_m^0}+
  \bra{\partial_b u_n^0}\partial_c u_m^0\rangle
  \right)
\right].
\eeq
Interchanging $b\leftrightarrow c$ in the second term and combining
with the first yields minus the third term, which concludes the proof.

\section*{References}

\bibliographystyle{iopart-num}
\bibliography{paper}

\end{document}